\documentclass[10pt]{article}
\usepackage{amssymb}
\def\be{\begin{equation}}
\def\ee{\end{equation}}
\def\bea{\begin{eqnarray}}
\def\eea{\end{eqnarray}}
\def\re#1{(\ref{#1})}

\begin{document}
%\rightline{initial date : Mai 4, 2010}             %%%
%\rightline{printed : \today}
%\rightline{file name: sufficiencyp.tex}
% \rightline{hep-th/yymmnnn}

\vskip 1. cm
\begin{center}\begin{Large}

{\bf Necessary conditions for Ternary Algebras}  
\end{Large}
\\
\vspace{0.5cm}

{David B. Fairlie$^1$, Jean Nuyts$^2$}
\vskip 0.6 true cm
{\small  
$^1$ david.fairlie@durham.ac.uk\\
{\it  Department of Mathematical Sciences, University of Durham, \\[1pt]
Science Laboratories, South Rd, Durham DH1 3LE, England}
\\[3pt]
$^2$  jean.nuyts@umons.ac.be\\                    %%%
{\it    Physique Th\'eorique et Math\'ematique,
Universit\'e de Mons,
\\[1pt]
 20 Place du Parc, B-7000 Mons, Belgium}
\\[3pt]
 }

\end{center}
\vspace{0.2cm}

\begin{abstract}
Ternary algebras, constructed from ternary commutators, or as we call them, ternutators,
defined as the alternating sum of products of three operators, have been shown to satisfy
cubic identities as necessary conditions for their existence. 
Here we examine the situation where                                          
we permit identities not solely constructed from ternutators or nested ternutators and we find 
that in general, these impose additional restrictions; for example, 
the anti-commutators or commutators of the operators                        
must obey some linear relations among themselves. 
\end{abstract}

\section{Introduction}

The subject of ternary algebras, a special case of n-Lie algebras is generally attributed to Fillipov
\cite{Fillipov}, but Filippov was following up on earlier
studies that had appeared in the mathematics literature, primarily by Kurosh
\cite{Kurosh} (as remarked in \cite{Zachos}).
Its appearance in Physics was due to the pioneering work of Nambu \cite{Nambu}, 
and more recently, the work of Bagger and Lambert \cite{Bagger} renewed interest 
in ternary algebras in the Theoretical Physics community (see also the review article [6]).
In general, a Ternary Bracket is a composition law for three operators, 
which is completely antisymmetric in the three operators; 
as for example Nambu Brackets, 
which are an extension of the idea of Poisson Brackets to three
functions. In this article we study the purely algebraic structure 
of the algebra, with a product structure for the operators clearly in mind, 
and we refer to such brackets as ternutators.

\section{Ternutator basics}

The ternutator bracket is a completely antisymmetrised 
trilinear composition law for three associative operators, 
just as the commutator is for two operators
\bea
\left[A,B,C\right]&\equiv& ABC+BCA+CAB-ACB-CBA-BAC\\ \label{com3}
                   &\equiv&        
                   \frac{1}{2}\left([[A,B],C]_{+}+[[B,C],A]_{+}+[[C,A],B]_{+}\right). 
                   \label{com3a}
\eea

Note the appearance of anti-commutators; 
if instead all the brackets in \re{com3} are commutators,  
the right hand side is the Jacobi identity for Lie Brackets, and is zero.
Corresponding operator algebras would read
\be
\left[A_i,A_j,A_k\right]=f_{ijk}^{\phantom{ijk}m}A_m,
\label{Lie3}
\ee
where the structure constants $f_{ijk}^{\phantom{ijk}m}$ 
are completely antisymmetric in $i,j,k$.

For a simplified notation let us write
\bea
(ijkl\dots)&\equiv& A_iA_jA_kA_l\dots
    \label{not1}\\
\left[ij\right]_{\pm}&\equiv&\left[A_i,A_j\right]_{\pm} 
    \label{not2}\\
\left[ijk\right]&\equiv&\left[A_i,A_j,A_k\right] 
    \label{not3}
\eea    
respectively
for the product of $n$ arbitrary operators \re{not1}, for the
anticommutator or commutator \re{not2}, and for the ternutator \re{not3}.
\section{Normal and non-normal order}

We associate to each operator a label $L$ equal to its index $L(A_{j_{_1}})=j_{_1}$. 
For a product of three operators $A_{j_{_1}}A_{j_{_2}}A_{j_{_3}}$, we say that they are in normal order 
if at least one of the set of indices
$\{j_1,j_{2}\}$ and $\{j_{2},j_{3}\}$ is in
increasing order. 
For three given operators, five of their products are in "`normal"' order and one is in
"non-normal" order. Example 
\bea
(321)&\quad&{\rm{non\!{-}normal\ order}}
    \label{nonorder000}\\
(123),(132),(213),(231),(312)&\quad&{\rm{normal\ order.}}    
\label{order000}
\eea
One clearly has
\be
(321)\equiv \left[321\right]+(123)-(132)-(213)+(231)+(312).
\label{basic}
\ee 
In other words, 
the triple non-normal product \re{nonorder000} 
can be written as a sum of normal triple products up
to a ternutator which 
is, through \re{Lie3}, 
of degree one in the operators (a decrease by two degrees).
 More generally, any product of three operators in non-normal order 
can be written as the sum of operators in normal order
up to a ternutator.      
\bea
&{\rm{For\ }}i_3>i_2>i_1&
    \nonumber\\
&(i_3i_2i_1)\equiv -\left[i_1i_2i_3\right]+(i_1i_2i_3)+(i_2i_3i_1)+(i_3i_1i_2)-(i_1i_3i_2)-(i_2i_1i_3).&           
\label{basicgen}
\eea  

To simplify, we consider                                   
products which involve only operators with different indices
\be
A_{j_{_1}}A_{j_{_2}}A_{j_{_3}}\dots \quad {\rm{with}}\quad A_{j_{_k}}\neq A_{j_{_m}}. \quad {\rm{for}}\quad k\neq m.
\label{different}
\ee

At the next ternutator       
level we have to define the non-normal product of one ternutator and two operators 
or of two ternutators and one operator leading to the appearance of a ternutator of ternutators. 
At higher levels,    
one obtains nested ternutators of ternutators.

Let us associate as follows 
a label $L$ with a nested ternutator. Take all the indices $i_1,\dots,i_n$ of the operators 
which enter in the nested ternutator and define     
\be
L=\min\{i_1,\dots,i_n\}.
\label{labeldef}
\ee

It is then easy to define the non-normal product of three nested ternutators 
$T_3,T_2$ and $T_1$ with label $L_3,L_2,L_1$ respectively.
They are those with $L_3>L_2>L_1$.
They are transformed into normal products by the obvious
\begin{equation}
T_3T_2T_1=-\left[T_1,T_2,T_3\right] + T_3T_1T_2 + T_1T_2T_3 +T_2T_3T_1 + T_3T_1T_2 - T_2T_1T_3 - T_1T_3T_2.\label{normalgen}
\end{equation}
The ternutator $\left[T_1,T_2,T_3\right]$ of higher nesting has label $L=L_1$.

Using these definitions, any product of different operators can be transformed in a sum of normal products.

Starting from a given product of operators, 
there are often many different paths which
can be followed to transform them in normal order. The difference between two results when they are different
lead what we call ternutator identities.

\section{General considerations about identities}
 
It is known that ternutators enjoy the seven Bremner-Nuyts identities among seven operators. These identities, \cite{Bremner}\cite{Nuyts}, play the r\^ole of the Jacobi identity for ternary algebras and  generate cubic necessary conditions on the structure constants of these algebras. These identities are also satisfied by Nambu Brackets,\cite{Nambu} a trilinear antisymmetric composition law for three operators, which associates with three functions $f(x,y,z),\ g(x,y,z),\ h(x,y,z)$ a ternary bracket of the form
 \begin{equation}\left[f,g,h\right] =\det\left|\begin{array}{ccc}
 \frac{\partial f}{\partial x}&\frac{\partial g}{\partial x}&\frac{\partial h}{\partial x}\\
 \frac{\partial f}{\partial y}&\frac{\partial g}{\partial y}&\frac{\partial h}{\partial y}\\
 \frac{\partial f}{\partial z}&\frac{\partial g}{\partial z}&\frac{\partial h}{\partial z} 
\end{array}\right|,\label{nambu}\end{equation}
just as the Poisson Bracket of two functions obeys the Jacobi identity of Lie brackets. This is discussed further in \cite{Curtright1}.
They are also satisfied by other trilinear composition laws, such as that of Awata et al \cite{Awata};
\begin{equation}
\left[A_i,A_j,A_k\right]_{Aw}                       
=\left[A_i,A_j\right]<A_k>+\left[A_j,A_k\right]<A_i>+\left[A_k,A_i\right]<A_j>,
\label{awata}
\end{equation}
where $<A_k>$ denotes the trace of the operator $A_k$. 
An important question is related to the sufficiency of these conditions. 
For Ternary algebras depending upon a composition law which 
intrinsically requires the composition of three operators, 
only ternutators of ternutators are allowed in the search for identities. 
However, since we also have a product at our disposal 
in terms of our definition \re{com3}, we can search 
for more general identical relations among the operators of the algebra. 
In this article we show that, much to our surprise, 
there exist identities involving four and six operators 
which lead also to necessary requirements. 
If we assume  linear independence of the anti-commutators  
(or commutators)
of the operators of the algebra, 
the identities for four (or six) operators severely 
restrict the allowable form of the structure constants.  
Conversely, the four operator and six operator identities 
may be interpreted as  linear relations among the anti-commutators 
or commutators of operators    
of the algebra.

\section{Products of Operators. Identities}

In this section, we discuss the identities which can be obtained 
starting with products of a certain number of operators. 

Quite generally, we suppose                 
that the basic operators conveniently labeled $A_1,A_2,\dots$  
(in some fixed order of the indices)
are defined to be linearly independent.

The identities among products of operators are of two kinds. 

\begin{enumerate}

\item
First, we have           
the sui generis identities which involve nested products
of termutators where all the operators in the products appear
in the form of ternutators only. It implies in particular that 
the products must involve an odd number of generators. In an earlier article 
\cite{Nuyts}, we showed that there are no sui generis identities for products 
of five operators and seven identities for products of seven operators.
For ternutator algebras \re{Lie3}, since the operators 
themselves are independent, they lead to cubic necessary conditions for
the structure constants.

\item
Identities where some operators may appear in 
an unnested positions.    
In this section,
we concentrate on such identities 
for products of four, five and six operators. 
There are obviously no such identities for products of   
two or three operators.    
\end{enumerate}
\subsection{Products of four operators}
The non-normal products of four operators are of three forms 
\begin{itemize}
\item
The exceptional product
\be
(4321)
\label{form1}
\ee
where both the three first operators and the three last operators 
are not in normal form.
\item
The products
\bea
&&({\underline{431}}2)\ ,\ ({\underline{421}}3)\ ,\ ({\underline{321}}4)
     \label{form2}\\
&&(1{\underline{432}})\ ,\ (2{\underline{431}})\ ,\ (3{\underline{421}}).   
     \label{form3}
\eea     
In the sets, we have underlined the operators which are not in normal form. They are the three first operators
in \re{form2} and the three last operators in \re{form3}. 
 
\end{itemize}

It is easy to see that,
using \re{basic}, the products in 
\re{form2} and in \re{form3} can be brought to normal form by following a path which is unique. 
This is not the case for the product $(4321)$ in \re{form1} where two different paths can be followed
depending on which set of three operators is used first the set $(432)$ or the set $(321)$. 
Let us follow the two paths.    
For path 1, one finds
\bea
({\underline{432}}1)&=&(\left[432\right]1)+(4231)+(3{\underline{421}})+(3241)-(2341)-(2{\underline{431}})
     \nonumber\\
     &=&\dots
           \nonumber\\
    &=&(\left[432\right]1)+(3\left[421\right])-(2\left[431\right])-(\left[321\right]4)
            \nonumber\\
       &&   +(4231)+(3412)-(3142)-(2413)
            \nonumber\\
       &&+(2143) -(1234)+(1324).
        \label{path1}
\eea                 
Following the same pattern, one finds for path 2
\bea
(4{\underline{321}})&=&
           (4\left[321\right])-(\left[421\right]3)+(\left[431\right]2)-(1\left[432\right])
            \nonumber\\ 
       &&   +(4231)+(3412)-(3142)-(2413)
            \nonumber\\
       &&+(2143) -(1234)+(1324).
        \label{path2}
\eea                
Subtracting the results of \re{path1} and \re{path2}, one finds the identity
\be
          I_4(4,3,2,1)=
           \left[\left[432\right]1\right]_{+}
          -\left[\left[431\right]2\right]_{+}
          +\left[\left[421\right]3\right]_{+}
          -\left[\left[321\right]4\right]_{+}
=0.
\label{fourid}
\ee   
Note the appearance of anti-commutators. This identity has also appeared in
Curtright and Zachos \cite{Curtright1}, in equation (84) of that paper,  but these authors did not pursue its implications. 
The identity can be written as
\be
I_4(i_4,i_3,i_2,i_1)=
 \sum_{ {\{\mu_1,\mu_2,\mu_3,\mu_4\}\in S_4 \atop  
      {\mu_1<\mu_2<\mu_3}}}  \hspace{-1 cm}         
    {\rm{sign}}(P)\left[\left[A_{\mu_1},A_{\mu_2},A_{\mu_3}\right],A_{\mu_4}\right]_{+}=0,
\label{Ifoursymbol} 
\ee  
where $S_4$ are permutations of $\{i_1,i_2,i_3,i_4\}$, and sign($P$) their sign.

These relations imply new necessary conditions on ternutator algebras (see below).

Indeed there also exist analogues to the identity of degree 4 for Nambu Brackets, three of them to be precise. These follow simply from the following argument. Consider the expansion of the determinant
\begin{equation}\det\left|\begin{array}{cccc}
\frac{\partial f}{\partial \alpha}&\frac{\partial g}{\partial \alpha}&\frac{\partial h}{\partial \alpha}&\frac{\partial k}{\partial \alpha}\\
 \frac{\partial f}{\partial x}&\frac{\partial g}{\partial x}&\frac{\partial h}{\partial x}&\frac{\partial k}{\partial x}\\
 \frac{\partial f}{\partial y}&\frac{\partial g}{\partial y}&\frac{\partial h}{\partial y}&\frac{\partial k}{\partial y}\\
 \frac{\partial f}{\partial z}&\frac{\partial g}{\partial z}&\frac{\partial h}{\partial z}&\frac{\partial k}{\partial z}
\end{array}\right|\equiv 0,\label{nambu2}\end{equation}
for $\alpha =x,\ y,$ or $z$.
Expanding the determinant on the first row we have
\begin{equation}
  \frac{\partial f}{\partial \alpha}\bigl[g,h,k\bigr]_{_{NB}}
 -\frac{\partial g}{\partial \alpha}\bigl[h,k,f\bigr]_{_{NB}}
 +\frac{\partial h}{\partial \alpha}\bigl[k,j,g\bigr]_{_{NB}}
 -\frac{\partial k}{\partial \alpha}\bigl[f,g,h\bigr]_{_{NB}}=0.
 \label{nambu3}
\end{equation}
These equations, for the various choices of $\alpha$, are the analogues of the identity \re{fourid}, for Nambu Brackets.

\subsection{Products of five operators}

Take five independent operators $A_1,\dots,A_5$, a REDUCE computation shows 
explicitly that there are ten 
independent identities involving the product of two of the operators and  one ternutator 
(made of the three remaining
operators), namely $[A_i,A_j,A_k]A_lA_m,\ A_l[A_i,A_j,A_k]A_m$ and $ A_lA_m[A_i,A_j,A_k]$. 
There are a priori 60 such products and hence 60 arbitrary coefficients; (5 choices for $l$,and 4 for
$m$ in each of the three classes). Grouping these terms  as  
$A_lA_m[A_i,A_j,A_k]+A_l[A_i,A_j,A_k]A_m$     
and $[A_i,A_j,A_k]A_lA_m+ A_l[A_i,A_j,A_k]A_m$, 
we see that they fall into ten sets which are identities 
in consequence of the four operator identity. 

There are 50 independent relations among these coefficients.   
Through REDUCE, it has been  checked explicitly that a basis for the ten identities is given by
the ten products
{\small                                  
\bea
  A_5I_4(4,3,2,1),\ A_4I_4(5,3,2,1),\ A_3I_4(5,4,2,1),
  \ A_2I_4(5,4,3,1),\ A_1I_4(5,4,3,2)&&
      \nonumber\\
   I_4(4,3,2,1)A_5,\ I_4(5,3,2,1)A_4,\ I_4(5,4,2,1)A_3,
  \ I_4(5,4,3,1)A_2,\ I_4(5,4,3,2)A_1&&.           
 \label{fiveid}
\eea
}
They all involve  four operator identities. Hence, no new identity exists at this degree.
This has also been checked explicitly by hand.
Remember that there are no sui generis identities for five operators.  

\subsection{Products of six operators}

There are 21 identities involving the product of     
six operators $A_1,\dots,A_6$ and which are quadratic 
in ternutators. 

\begin{itemize}

\item There are $C_6^3=20$ identities which are simple consequences of the four identity. 
They are indexed by the choice of separating the six operators in two non overlapping sets 
$\{i_1<i_2<i_3\}$ and $\{i_4<i_5<i_6\}$. They are conveniently written
\bea
\sum_{ {\{j_4,j_5,j_6\}\in P_3(\{i_4,i_5,i_6\})} \atop 
      {j_4<j_5}}  \hspace{-1 cm}         
    {\rm{sign}}(P)\left[\left[\left[A_{i_1},A_{i_2},A_{i_3}\right],A_{j_4},A_{j_5}\right]A_{j_6}\right]_{+}\quad&&
      \nonumber\\
-\hspace{-1 cm} \sum_{ {\{j_1,j_2,j_3\}\in P_3(\{i_1,i_2,i_3\})} \atop 
      {j_1<j_2}}           
    {\rm{sign}}(P)\left[\left[\left[A_{i_4},A_{i_5},A_{i_6}\right],A_{j_1},A_{j_2}\right]A_{j_3}\right]_{+}
&=&0.
\label{Isixnormal}
\eea  

\item
  
The remaining identity can be written democratically as

\bea
&&
\hspace{-0.5cm}
\hskip-5pt
\lefteqn{6
\hskip-35pt
%\hskip-0pt
\sum_{{\mu\in S_6
\atop\mu_1<\mu_2<\mu_3,\;\mu_4<\mu_5<\mu_6,\;\mu_1<\mu_4}}
\hskip-40pt
\epsilon_{\mu_1\mu_2\mu_3\mu_4\mu_5\mu_6}\,
  \Bigl[\Bigl[A_{\mu_1},A_{\mu_2},A_{\mu_3}\Bigr],
  \Bigl[A_{\mu_4},A_{\mu_5},A_{\mu_6}\Bigr]\Bigr]_{_{-}}}
 \nonumber   \\
  &&
  \hspace{-0.5cm}
   -
  \hskip-22pt
  \sum_{{\mu\in S_6
  \atop\mu_1<\mu_2<\mu_3,\;\mu_4<\mu_5}}
%  \hskip-40pt
  \hskip-15pt
  \epsilon_{\mu_1\mu_2\mu_3\mu_4\mu_5\mu_6}\,
      \Bigl[\Bigl[\Bigl[A_{\mu_1},A_{\mu_2},A_{\mu_3}\Bigr],
      A_{\mu_4},A_{\mu_5}\Bigr],A_{\mu_6}\Bigr]_{_{-}}=0.
\label{id6}
\eea
This identity can be written in many apparently different forms adding terms which are zero 
when using the four-identity.

\end{itemize} 

We have first obtained the exceptional identity \re{id6} by using 
Reduce to built all the possible identities. 
Another approach to this identity is to use the Bremner-Nuyts \cite{Bremner}\cite{Nuyts} seven identity in the form of equation (13)
of \cite{Nuyts} singling out a $A_7$ operator on the right. 
Then the coefficient of $A_7$ constitutes the identity 
at level 6.
Another way is to use
the path analysis  for six operators,  one obtains an
    identity by comparing for the initial configuration $(654321)$,
    the path starting with
    $(\,{\underline{6\,5\,4}}\,\ {\underline{3\,2\,1}}\,)$ with the path
    starting for example by
    $(\,6\,{\underline{5\,4\,3}}\,2\,1\,)$.
    Any other path starting  with
    $(\,6\,5\,{\underline{4\,3\,2}}\,1\,)$ or
    $(\,6\,5\,4\,{\underline{3\,2\,1}}\,)$ and where $6$ is a spectator
    will give the same identity as all these three paths are equivalent
    as they involve five operators known to be equivalent through I4. 
    This is another proof that the 6-identity is unique.

There seem to be no identities for six operators linear in the structure constants, i.e. of first degree 
in the ternutators of the basic operators, except those which follow from the 
4-identity. This was essentially proved using Reduce.

% Using the path analysis  for six operators,  one obtains an
%     identity by comparing for the initial configuration $(654321)$,
%     the path starting with
%    $(\,{\underline{6\,5\,4}}\,\ {\underline{3\,2\,1}}\,)$ with the path
%    starting for example by
%    $(\,6\,{\underline{5\,4\,3}}\,2\,1\,)$.
%    Any other path starting  with
%    $(\,6\,5\,{\underline{4\,3\,2}}\,1\,)$ or
%    $(\,6\,5\,4\,{\underline{3\,2\,1}}\,)$ and where $6$ is a spectator
%    will give the same identity as all these three paths are equivalent
%    as they involve five operators.

%    Any other starting configuration
%    $(\bullet\bullet\bullet\bullet\bullet\bullet)$
%    which allows a starting path
%    $({\underline{\bullet\,\bullet\,\bullet}}\,\
%    {\underline{\bullet\,\bullet\,\bullet}})$ will give the same identity
%    as can easily be seen by relabeling the operators.
%    Hence there exists one and only one new genuine identity involving six
%    operators.

\section{{\bf{Conditions on the structure constants from the identities of degree four}}}

The identity \re{fourid} implies the following type of conditions on the structure constants
\be
f_{432}^{\phantom{ijk}m}[A_m,A_1]_{_{+}}
-f_{431}^{\phantom{ijk}m}[A_m,A_2]_{_{+}}
+f_{421}^{\phantom{ijk}m}[A_m,A_3]_{_{+}}
-f_{321}^{\phantom{ijk}m}[A_m,A_4]_{_{+}}=0.
\label{fouridcond}
\ee
If we assume that the anti-commutators are linearly independent, we find
the conditions
\begin{itemize}
\item 
\be
{\hskip - 0.7 cm}{\rm{From\ }}A_1^2\quad\longrightarrow\quad f_{432}^{\phantom{ijk}1}=0.
\label{a1square}
\ee
\item
\be
\quad{\rm{From\ }}A_1A_2\quad\longrightarrow\quad f_{432}^{\phantom{ijk}2}-f_{431}^{\phantom{ijk}1}=0
\label{a1a2}
\ee
\item
Using $m\notin \{1,2,3,4\}$, we recover a condition of the form \re{a1square}
\be
{\rm{From\ }}A_1A_5\quad\longrightarrow\quad f_{432}^{\phantom{ijk}5}=0.
\label{a1a3}
\ee

\end{itemize}
Summarizing the results, we find
\bea
f_{ijk}^{\phantom{ijk}m}&=&0\quad {\rm{for\ }} m\notin \{i,j,k\}
     \nonumber\\
f_{ijm}^{\phantom{ijk}m} &=&f_{ijk}^{\phantom{ijk}k}\  \forall\  m,k\notin\{i,j\}.     
\label{struccond4}
\eea
These results severely restrict the possible ternary algebras unless  we drop the assumption of 
independence, and re-interpret \re{struccond4} 
as a set of linear relations among the anti-commutators of the operators. This is a surprising result.

\section{Alternative iterative approach}

There is another way of looking at the identities which we have found.
Consider the following iterative situation;
\bea \left[[A_1,A_2],A_3\right]+\left[[A_2,A_3],A_1\right]+\left[[A_3,A_2],A_1\right] &=& 0,\label{alt1}\\
  \left[[A_1,A_2],A_3\right]_{+}+ \left[[A_2,A_3],A_1\right]_{+}+\left[[A_3,A_2],A_1\right]_{+} &=&     2[A_1,A_2,A_3].
  \label{alt2}
\eea
The first is the Jacobi identity, the second is twice the ternutator $[A_1,A_2,A_3]$
 Now add one more operator
\begin{eqnarray}
\left[[A_1,A_2,A_3],A_4\right]_{+} -\left[[A_2,A_3,A_4],A_1\right]_{+}&&\nonumber\\
+\left[[A_3,A_4,A_1],A_2\right]_{+}-\left[[A_4,A_1,A_2],A_3\right]_{+}&=&0.\label{alt3}
\end{eqnarray}
 \begin{eqnarray}\left[[A_1,A_2,A_3],A_4\right]-\left[[A_2,A_3,A_4],A_1\right]&&\nonumber\\
+\left[[A_3,A_4,A_1],A_2\right]-\left[[A_4,A_1,A_2],A_3\right]&=&2\left[A_1,A_2,A_3,A_4\right].
\label{alt4}
\end{eqnarray}

The first equation is the 4-identity $I_4(4321)=0$ and the second equation
is the definition of the antisymmetric 4-bracket $[A_1,A_2,A_3,A_4]$
with an additional factor of 2. 

This pattern of repeated nested alternating commutators and anti-commu\-tators
persists.
Writing the antisymmetric $n$-bracket in a generalized notation \re{not2}, \re{not3} as
\be
    \left[i_1i_2\dots i_n\right]=\left[A_{i_1},A_{i_2},\dots,A_{i_n}\right],   
\label{not4}
\ee    
we find 
\begin{itemize}

\item
For an even number of operators
\bea
\left[i_1i_2\dots i_{2n}\right] 
&=&\frac{1}{2}\sum_{\rm {cyclic}}{\rm{sign}} (C)\left[\left[i_1i_2\dots i_{2n-1}\right]i_{2n}\right]_{_{-}} 
   \nonumber\\
0
&=&\sum_{\rm{cyclic}} {\rm{sign}} (C)
\left[\left[i_1 i_2 \dots i_{2n-1}\right]i_{2n}\right]_{_{+}}
\label{altereven}
\eea
where the summation is over the cyclic permutation of $\{i_1,i_2,\dots, i_{2n}\}$
and sign$(C)$ is the sign of the permutation.
\item
For an odd number of operators
\bea
0
&=&\sum_{\rm{ cyclic}}{\rm{sign}} (C)\left[\left[i_1i_2\dots i_{2n}\right] i_{2n+1}\right]_{_{-}} 
   \nonumber\\
   \left[i_1i_2\dots i_{2n+1}\right] 
&=&\frac{1}{2}\sum_{\rm {cyclic}}{\rm{sign}} (C) \left[\left[i_1i_2\dots i_{2n}\right] i_{2n+1}\right]_{_{+}} 
\label{alterodd}
\eea
with summation over the cyclic permutations of $\{i_1,i_2,\dots, i_{2n+1}\}$.
\end{itemize}

As has been demonstrated, the identity  at level 5 is not new. The level 6 identities 
contain the 20 identities \re{Isixnormal} in a democratic fashion. 

\section{Discussion and Conclusion}

For a pure mathematician, the only relevant identities 
are those composed of only ternary operations, i.e operations
sui generis, of nested ternutators, just as for Lie algebras, 
the relevant identities are those composed of iterated or nested commutators, 
and SuperJacobi identities are ignored. 
It is known that in this case 
the Jacobi identity is both necessary and sufficient. 
In the case  of  ternary algebras, 
we have demonstrated an identity at the level of four operators
(and one identity at the level of six operators),  
which involves anti-commutators (or commutators)    
and which can be interpreted in various ways. 
Either we assume that anti-commutators (or commutators) 
of all the operators involved 
are independent of one another, 
in which case there are very few viable ternary algebras, 
or we assert that having found a representation 
of the ternutator algebra, linear relations amongst the 
anti-commutators (or commutators)      
must be automatically satisfied. 
Perhaps we may exploit these relations in the search for representations of ternutators.

\end{document}